\documentclass[APS,twocolumn,showpacs,preprintnumbers,amsmath,amssymb]{revtex4}

\usepackage{graphicx}                               % include figure files
\usepackage{amsfonts,amssymb, amsmath}
\usepackage[amsmath,thmmarks]{ntheorem} 						% [thmmarks,standard,thref]
\usepackage{quantsymb}                              % my own package
\usepackage{psfrag}                                 % to change some characters in a *.eps file
\usepackage{wasysym}																% various symbols (hexagon, ...)

\newtheorem{Res}{Result}

\theoremstyle{nonumberplain}
\theoremheaderfont{\normalfont\itshape}
\theorembodyfont{\normalfont}
\theoremseparator{.}
\theoremsymbol{\ensuremath{\square}}
\newtheorem{Pf}{Proof}

\newcommand{\ie}{\emph{i.e.\@} }

\newcommand{\etc}{etc.}

\newcommand{\quoded}{\emph{q.e.d.\@} }

\newcommand{\romone}{\text{I}}
\newcommand{\romtwo}{\text{II}}
\newcommand{\id}{\openone}

\newcommand{\WCE}{W}
\newcommand{\SCP}{S}

\DeclareMathOperator{\eig}{eig}
\DeclareMathOperator{\tr}{tr}

\begin{document}

\title{Entanglement Distribution in Pure-State Quantum Networks}
\author{S\'ebastien Perseguers}
\email{sebastien.perseguers@mpq.mpg.de}
\author{J. Ignacio Cirac}
\affiliation{Max-Planck--Institut f\"ur Quantenoptik, Hans-Kopfermann-Str. 1, D-85748 Garching, Germany}
\author{Antonio Ac\'in}
\author{Maciej Lewenstein}
\affiliation{ICFO-Institut de Ci\`encies Fot\`oniques, Mediterranean Technology Park, E-08860 Castelldefels (Barcelona), Spain\\
ICREA-Instituci\'o Catalana de Recerca i Estudis Avan\c cats, Lluis Companys 23, 08010 Barcelona, Spain}
\author{Jan Wehr}
\affiliation{Department of Mathematics, The University of Arizona, Tucson, AZ 85721-0089, USA}

\date{October 31, 2007}

% -----------------------------------------------------------------------------
% ABSTRACT
% -----------------------------------------------------------------------------
\begin{abstract}
We investigate entanglement distribution in pure-state quantum networks. We
consider the case when non-maximally entangled two-qubit pure states are shared
by neighboring nodes of the network. For a given pair of nodes, we investigate
how to generate the maximal entanglement between them by performing local
measurements, assisted by classical communication, on the other nodes. We find
optimal measurement protocols for both small and large 1D networks. Quite
surprisingly, we prove that Bell measurements are not always the optimal ones
to perform in such networks. We generalize then the results to simple small 2D
networks, finding again counter-intuitive optimal measurement
strategies. Finally, we consider large networks with hierarchical lattice
geometries and 2D networks. We prove that perfect entanglement can be
established on large distances with probability one in a finite number of steps,
provided the initial entanglement shared by neighboring nodes is large enough.
We discuss also various protocols of entanglement distribution in 2D networks
employing classical and quantum percolation strategies.
\end{abstract}

\pacs{03.67.-a,~03.67.Bg}				% The Physics and Astronomy Classification Scheme.
%\keywords{Suggested keywords}	% Use showkeys class option

\maketitle

% *****************************************************************************
% SECTIONS
% *****************************************************************************

% #############################################################################
% Introduction
% #############################################################################
\section{Introduction}
\label{sec:intro}
Quantum Networks \cite{CZKM97,BBM+07} play a key role in quantum information
processing. In such networks, quantum states can be prepared initially and
shared between neighboring nodes (or stations), \ie entanglement can be generated,
and this resource is then to be used for quantum communication
\cite{E91,BBC+93}, or distributed quantum computation \cite{CEHM99} involving
arbitrary nodes of the network. One of the main tasks is then to design
protocols that use the available quantum correlations to entangle two nodes of
the network, and to optimize these protocols in terms of final entanglement and
probability of success.

\begin{figure}
    \psfrag{a1}[][]{$\alpha_1$}	\psfrag{a2}[][]{$\alpha_2$}	\psfrag{an}[][]{$\alpha_{N+1}$}
    \psfrag{a}[][]{$\alpha$}		\psfrag{b}[][]{$\beta$}			\psfrag{c}[][]{$\gamma$}
    \psfrag{(a)}[][]{(a)}				\psfrag{(b)}[][]{(b)}				\psfrag{(c)}[][]{(c)}
    \psfrag{p}[][]{$\varphi_m$}
    \psfrag{M1}[][]{$M_1$}			\psfrag{M2}[][]{$M_2$}			\psfrag{Mn}[][]{$M_N$}	\psfrag{M}[][]{$M$}
    \psfrag{A}[][]{$A$}					\psfrag{B}[][]{$B$}					\psfrag{C}[][]{$C$}			\psfrag{D}[][]{$D$}
  \begin{center}
      \includegraphics[width=0.95\linewidth]{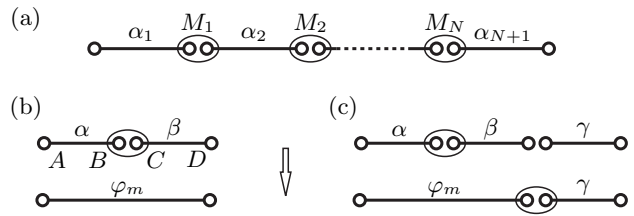}
      \caption{Notation and examples of 1D networks: (a) the standard quantum repeater scenario;
      	(b) entanglement swapping; (c) a two-repeater system.}
      \label{fig:notation1D}
  \end{center}
\end{figure}

A set of quantum repeater stations, for instance (see Fig.~\ref{fig:notation1D}a),
can be considered as a 1D quantum network, where the aim is to establish
quantum communication over large distances \cite{BDCZ98,DBCZ99,CTSL05,HKBD06}.
It is well known that the simple entanglement swapping \cite{ZZHE93} procedure
can achieve this goal, but (except for the unrealistic case of perfect
resources and operations) the probability of obtaining entanglement between the
end-nodes of such a network decays exponentially with the number of repeaters.
This problem can be overcome by the more sophisticated quantum repeaters
protocols \cite{BDCZ98,DBCZ99,CTSL05,HKBD06} which intersperse ``connection
steps'' (entanglement swapping) with purification steps and require only
polynomial decay, thus opening the way for feasible long-distance quantum
communication.

In two-dimensional or higher dimensional lattices of large size, a perfect connection between any two nodes
is possible with a probability that is strictly greater than zero, even with imperfect resources.
This can be  achieved by the so-called classical and quantum percolation strategies \cite{ACL07}, in which initially, or after some preparatory measurements, respectively,  one converts all bonds into singlets with a probability $p$.
This result is very encouraging, but remains of little use for finite and small networks \cite{KLH+04,EAM+05}.
The aim of the present paper  is twofold: first, we investigate and derive optimal local measurement protocols for
simple networks of finite size. In particular, we consider certain 1D and 2D networks
of nodes that consist of $z$
qubits, where $z$ is the number of neighbours.  Neighboring nodes share \emph{partially} entangled pure states.  We apply then local quantum operations to the nodes, assuming that these operations are noiseless. We first address the question of
optimal entanglement propagation, or in another words entanglement swapping, in small networks consisting of three or four
nodes only. The insights obtained for these simple situations are then used as
building blocks for larger 1D and 2D quantum networks, as well as networks with hierarchical geometry.

Our second aim is to discuss examples of hierarchical ``diamond'' and ``tree'' lattices
in which perfect entanglement on arbitrary large distances  can be achieved in a finite number of steps (measurements). Provided that sufficiently large but not necessarily maximal entanglement is present, we can convert connections along a given line into prefect singlets. Finally, we consider various kinds of percolation strategies: the one presented in Ref.~\cite{ACL07}, which employs a change in the lattice connectivity due to quantum measurements, and  a different one in a triangular lattice, where the optimal singlet conversion strategy is used. Both of these protocols essentially demonstrate that the quantum percolation thresholds are lower than their classical counterparts. Equally interesting we propose to use the optimal singlet conversion strategy to transform a square lattice into two independent square lattices of doubled size, for which the percolation probability is larger than in the original lattice.\\

% Outline
% -----------------------------------------------------------------------------
\paragraph*{Outline}
In Section II we fix the notation and define the figures of merit used for evaluating the
measurements efficiency: the concurrence (C), the so-called worst case entanglement (WCE) and the singlet conversion probability (SCP). In Sec.~\ref{sec:1d}  we describe the strategies maximizing these
quantities for some 1D  networks, starting from a simple one-repeater configuration, consisting of two bonds with two imperfectly entangled pairs on them. Interestingly, there exists a strategy that conserves the averaged singlet conversion probability \cite{BVK99}; the protocols however does not scale with the number of repeaters, as expected. %the prize to pay in this protocol is that the original two pure entangled states have to be replaced by a mixed state.
The second subsection of Sec.~\ref{sec:1d} deals with the problem of two repeaters, that is three bonds. Here the optimization of the SCP is much more complex: in some conditions we obtain that the optimal measurements do not correspond to a Bell measurement, \ie von Neumann measurements in a Bell basis of orthonormal maximally entangled states. This result is somewhat analogue to the recent result by Mod\l awska and Grudka \cite{MG07}, who have demonstrated that non-maximally entangled states can be better for the realization of multiple linear optical teleportation
in the scheme of Knill, Laflamme and Milburn \cite{KLM01}.  The last part of this section deals with large 1D network (\ie in the limit of infinite size network). Here we prove that the probability of establishing entanglement over large distances decays exponentially. We present optimal strategies for the concurrence and the WCE, and upper bounds for the SCP.

In Sec.~\ref{sec:2d} we turn to the simplest small network in 2D: a square. We obtain similar results as in the case of two repeaters in 1D, indicating that Bell measurements not always provide the best protocol. In Sec.~\ref{sec:hier} we apply the results of previous sections to networks of large size and hierarchical geometry, that is, lattices that iterate certain geometric structures, so that at each level of iterations the number of nodes, or the number of neighbors changes. We consider two kinds of hierarchical lattices: first we discuss the so-called ``diamond'' lattice, for which we prove that for sufficiently large initial entanglement, one can establish perfect entanglement on large scales (\ie some lower levels of iteration) in finite number of steps.
A somewhat simpler result holds for the simplest possible double Cayley tree lattice, in which in each step
of iteration each bond branches into two. For such lattices, if the initial entanglement is large enough, perfect entanglement can be established at each level of iteration.

Finally, in Sec.~\ref{sec:2D} we consider genuine 2D lattices. First, using similar method as in Sec.~\ref{sec:hier} we show that for a sufficiently broad strip of a square lattice, we can convert connections of a given line along the strip  into a line of perfect singlets, provided, of course, that initially an imperfect, but sufficiently large entanglement is present. Second we reconsider percolation strategies and discuss the example of hexagonal lattice with double bonds from
Ref.~\cite{ACL07}, and a triangular lattice with variable bonds. In the first of these examples quantum measurements lead to local reduction of the SCP, but change the geometry of the lattice, increasing its connectivity and thus the classical percolation threshold.
In the second example we use a protocol optimizing the SCP to transform the original lattice to a new one with the same geometry, but with a higher probability $p$ of getting a singlet on a bond. Similarly, we discuss a different type of strategy, where by using the optimal singlet conversion protocol we transform a square lattice into two independent square lattices with the same mean SCP as the initial one. We prove that the classical  probability of connecting a pair of neighboring points in the initial lattice (two neighboring point from the two lattices) to another such pair is strictly larger for the case of two lattices. We conclude then in   Sec.~\ref{sec:conclu}.

% #############################################################################
% Notations
% #############################################################################
\section{Preliminaries: Notation and Basic Notions}
\label{ssec:intro-notations}

% notation1D.eps ::::::::::::::::::::::::
% :::::::::::::::::::::::::::::::::::::

A pure state of two qubits is represented by a solid line in the figures and is
written (except when specified) as
\begin{equation}
    \ket{\alpha}=\sqrt{\alpha_0}\ket{00}+\sqrt{\alpha_1}\ket{11},
    \label{eqn:stateAB}
\end{equation}
where $\alpha_0+\alpha_1=1$ and $\alpha_0\geq\alpha_1$ (it is assumed that local
basis rotations are performed whenever necessary for the states to be written in
that way). This defines the Schmidt decomposition of a pure state of two qubits,
while $\alpha_0$ and $\alpha_1$ are their Schmidt coefficients.

% -----------------------------------------------------------------------------
% Entanglement Swapping
% -----------------------------------------------------------------------------
\subsubsection{Entanglement Swapping}
\label{sssec:intro-notations-swap}
A basic operation for propagating entanglement over larger distances is the so
called ``entanglement swapping'', see Fig.~\ref{fig:notation1D}b. A POVM
is described by $n$ positive operators $E_m\in M_4(\mathbb{C})$
satisfying the completeness relation $\sum_{m=1}^n E_m^{}=\id_4$. Let
\[\rho_m=\tr_{BC}\big((\id_2\otimes E_m\otimes\id_2)\,\ketbra{\alpha\beta}{\alpha\beta} \big)\]
be the resulting state of the measurement $M$, which occurs with a probability $p_m=\tr(\rho_m)$.
We consider in that paper projective measurements only, \ie $n=4$ and $E_m=\ketbratext{u_m}{u_m}$
for some normalized state $\ket{u_m}$. For such measurements the smallest Schmidt coefficient of $\rho_m$
is $\lambda_m=\min\{\eig(\widetilde{\rho}_m)\}/p_m$,
where $\widetilde{\rho}_m=\tr_A(\rho_m)$, or equivalently
\begin{equation}
    \lambda_m = \frac{1}{2}\bigg(1-\sqrt{1-\frac{4 \det(\widetilde{\rho}_m)}{p_m^2}}\bigg).
    \label{eqn:schmidtFromDet}
\end{equation}
Considering the following map from
$\mathbb{C}^2\otimes\mathbb{C}^2$ to $M(\mathbb{C},2)$:
\begin{equation}
    \ket{a}=\sum_{i,j=0}^1 a_{ij}\ket{ij}\quad\mapsto\quad\widehat{a}=
    \begin{pmatrix}a_{00}&a_{01}\\a_{10}&a_{11}\end{pmatrix},
\end{equation}
one can show that $\widetilde{\rho}_m$ is now equal to $X_m^{}X_m^{\dagger}$, with
$X_m = \widehat{\alpha}\,\widehat{u}_m\,\widehat{\beta}$.\\
The concurrence of a state $\varphi$ is by definition $C(\varphi) \equiv 2\,
|\det(\widehat{\varphi})|$. Therefore, the concurrence $C_m$, the smallest Schmidt
coefficient $\lambda_m$ and the outcome probability $p_m$ are explicitly
given by
\begin{subequations}
\begin{align}
    C_m	&= \frac{2\,|\det(X_m)|}{p_m}
         = \frac{\sqrt{\alpha_0\alpha_1\beta_0\beta_1}}{p_m}\,C(u_m),
        \label{eqn:projM-C}\\
    \lambda_m &= \frac{1}{2}\left(1-\sqrt{1-C_m^2}\right),
        \label{eqn:projM-lambda}\\
    p_m &= \sum_{i,j=0}^1\alpha_i\beta_j\,|\widehat{u}_{m,ij}|^2.
        \label{eqn:projM-p}
\end{align}
\label{eqn:projM}
\end{subequations}

We now turn to Bell measurements. Starting from the computational basis $\{\ket{0},\ket{1}\}$ of a single qubit,
we define the new orthonormal basis $\{\ket{\arrowU},\ket{\arrowD}\}$
\begin{equation}
    \begin{pmatrix}\ket{\arrowU}\\\ket{\arrowD}\end{pmatrix}=U
    \begin{pmatrix}\ket{0}\\\ket{1}\end{pmatrix},\quad U\in\mathcal{U}(2),
\end{equation}
and the Bell vectors
\begin{equation}
    \ket{\Phi^{\pm}}=\frac{\ket{\arrowUU}\pm\ket{\arrowDD}}{\sqrt{2}}
    \quad\text{and}\quad
    \ket{\Psi^{\pm}}=\frac{\ket{\arrowUD}\pm\ket{\arrowDU}}{\sqrt{2}}.
\end{equation}
Two specific bases play a key role in this paper:
the computational or ``ZZ'' basis, where the vectors $\ket{\arrowU}$    and $\ket{\arrowD}$
for both qubits are the eigenvectors of the Pauli matrix $\sigma_z$,
and the ``XZ'' basis, where the first basis is chosen as being the eigenvectors of $\sigma_x$.
Although we could in principle parameterize the Bell states in that way,
calculations are much easier and clearer in the ``magic basis'' defined as
\cite{HW97}
\begin{equation}
    (\widehat{\Phi}_1,\,\widehat{\Phi}_2,\,\widehat{\Phi}_3,\,\widehat{\Phi}_4) =
    (\id_2,\,-i \sigma_z,\,i \sigma_y,\,-i \sigma_x)\ket{\Phi^+},
\end{equation}
which is nothing but the usual Bell basis with some specific phases.
In this basis, the concurrence of a state $\ket{\mu}=\sum_{i=1}^4\mu_i\ket{\Phi_i}$
simply reads $C(\mu) = \big|\sum_{i=1}^4 \mu_i^2\big|$.
It follows that the coefficients $\mu_i$ of a Bell state (whose concurrence is 1 by definition)
have all the same phase; hence we can choose them as being real.
Let a set of four such states $\{\mu_m\}$, so that the matrix
$(\mu_{m,i})$ belongs to $\mathcal{SO}(4)$. Then the probabilities given in Eq.~(\ref{eqn:projM-p})
read
\begin{equation}
    p_m = p_{\min} \left(\mu_{m,1}^2+\mu_{m,2}^2\right) +
                p_{\max} \left(\mu_{m,3}^2+\mu_{m,4}^2\right),
    \label{eqn:notation-p}
\end{equation}
with
\begin{equation}
    p_{\min} = \frac{\alpha_0 \beta_1+\alpha_1 \beta_0}{2}
        \quad\text{and}\quad
    p_{\max} = \frac{\alpha_0 \beta_0+\alpha_1 \beta_1}{2}.
    \label{eqn:notation-p-interval}
\end{equation}

We emphasize the fact that (given two states $\alpha$ and $\beta$), the outcome probabilities completely characterize
a Bell measurement, since $\lambda_m$ depends only on $p_m$ for $C(u_m)=1$,
see Eq.~(\ref{eqn:projM}).

% -----------------------------------------------------------------------------
% Figures of Merit
% -----------------------------------------------------------------------------
\subsubsection{Figures of Merit}
\label{sssec:intro-merit}

We describe here three figures of merit used to evaluate the usefulness of an entanglement swapping protocol:
the concurrence, the singlet conversion probability (SCP) and the worst-case entanglement (WCE).
All these figures of merit take value in the interval $[0,1]$.\\

% C
% -----------------------------------------------------------------------------
\paragraph*{Concurrence}
The average concurrence of a measurement $M$ is defined as
\begin{equation}
    C_M=\sum_m p_m\,C_m,
    \label{eqn:merit-C}
\end{equation}
where $C_m$  is the concurrence of the outcome $m$.\\

% Worst-Case Entanglement
% -----------------------------------------------------------------------------
\paragraph*{WCE}
The idea of the WCE is to find a
measurement optimizing the entanglement for all its outcomes. Taking the
smallest Schmidt coefficient as the entanglement measure we define the WCE as
\begin{equation}
    \WCE_{M}=2\,\underset{m}{\min}\{\lambda_m\}.
\end{equation}

% SCP
% -----------------------------------------------------------------------------
\paragraph*{SCP}
We consider here the probability of conversion of a given state into a
perfect singlet. A result of majorization theory \cite{V99,NV01} tells us
that a state $\ket{\alpha}$ (\ref{eqn:stateAB}) can be converted into a singlet by LOCC with maximal
probability $2\,\alpha_1$; this is the ``Procrustean method'' of entanglement concentration
described in \cite{BBPS96}. We define the average SCP for a measurement $M$ as
\begin{equation}
    S_{M}=2\sum_m p_m\,\lambda_m,
\end{equation}
where $\lambda_m$ is the smallest Schmidt coefficient of the outcome $m$. Since
this figure of merit is used for different systems, we sometimes use
the following notation for clarity:
\[
    \SCP_M^{(N)}(\alpha_{1,0},\,\alpha_{2,0},\,\ldots,\,\alpha_{N+1,0}),
\]
where $N$ means the number of repeaters of a 1D chain consisting of $N+1$ states
$\alpha_1,\,\alpha_2,\,\ldots,\,\alpha_{N+1}$, as depicted in Fig.~\ref{fig:notation1D}a.

% #############################################################################
% 1D
% #############################################################################
\section{1D Networks}
\label{sec:1d}
Before studying the two-dimensional networks, it is worth looking at
systems made of one or two repeaters only. In fact, some interesting properties of
these small systems can then be used in more elaborated strategies for larger
networks. For instance, the important fact that the SCP does not decrease after
one measurement (\S\ref{sssec:1d-1r-scp}) allows one to get better results for the percolation on
honeycomb lattices \cite{ACL07}. Another important and surprising result
is that Bell measurements are not, in general, the measurements that maximize the
SCP (\S\ref{sssec:1d-2r-scp}), although they are the best ones for the average
concurrence and the WCE. Previous results on 1D networks can also be found in Refs.~\cite{BVK98,HS00}.

% =============================================================================
% One Repeater
% =============================================================================
\subsection{One Repeater}
\label{ssec:1d-1r}
We consider in this section a system consisting of two states $\alpha$ and $\beta$
joined by a single repeater, see Fig.~\ref{fig:notation1D}b.
We first prove a general statement on Bell measurements, and then
describe the measurements that maximize our three figures of merit.

% -----------------------------------------------------------------------------
% Bell Measurements
% -----------------------------------------------------------------------------
\subsubsection{Bell Measurements and Outcome Probabilities}
\label{sssec:1d-1r-bell}
The following result is very useful when trying to maximize the SCP
over the set of Bell measurements (the proof is given in App.~\ref{sec:app-proof-result}):\\

% Result
% .............................................................................
\begin{Res}\label{res:bell-prob}
    Outcome probabilities for a one-repeater Bell measurement
  \begin{quote}
    Let $\{x_m\}$ be four real numbers that add up to one and that lie in
    the interval $[p_{\min},\,p_{\max}]$. Then there exists a Bell measurement
    whose outcome probabilities $p_m$ are equal to $x_m$.
  \end{quote}
\end{Res}

% -----------------------------------------------------------------------------
% C and WCE
% -----------------------------------------------------------------------------
\subsubsection{Maximizing the Concurrence and the WCE}
\label{sssec:1d-1r-cwce}
It is clear from Eqs.~(\ref{eqn:projM-C}, \ref{eqn:merit-C}) that any Bell measurement,
\ie $C(u_m)=1\;\forall\,m$, maximizes the average concurrence, and therefore
\begin{equation}
    C_{\max}=2\sqrt{\alpha_0\alpha_1\beta_0\beta_1}.
\end{equation}
The result of the maximization of the WCE is summarized in the following result:\\

% Result
% .............................................................................
\begin{Res}\label{res:xz-basis}
    Best WCE strategy for one repeater
  \begin{quote}
    The maximum value of $\WCE$ for a one-repeater system is reached by the Bell
    measurement in the XZ basis, with
    \begin{equation}
            \WCE_{\max}=\WCE_{\text{XZ}}=1-\sqrt{1-16\,\alpha_0\alpha_1\beta_0\beta_1}.
        \end{equation}
        \end{quote}
\end{Res}

% Proof
% .............................................................................
\begin{Pf}  \textit{(By contradiction).}
The Bell states $\ket{u_m}$ in the XZ basis are given by the columns of the matrix
\[
    M_{\text{XZ}} = \frac{1}{2}
    \begin{pmatrix}
        -1 & 1 & 1 & 1 \\ 1 & -1 & 1 & 1 \\ 1 & 1 & -1 & 1 \\ 1 & 1 & 1 & -1
    \end{pmatrix},
\]
hence $p_m=1/4$ and $2\lambda_m=1-\sqrt{1-16\,\alpha_0\alpha_1\beta_0\beta_1}
\;\forall\,m$. Now suppose that there exists a measurement $M$ described by the set
$\left\{E_m=\ketbratext{u_m}{u_m}\right\}_{m=1}^n$, with $n\geq4$,
such that $\WCE_{M}>\WCE_{\text{XZ}}$. Then each $\lambda_m$ has to be strictly greater
than the smallest Schmidt coefficient of the outcomes in the XZ basis.
Thus, from Eq.~(\ref{eqn:schmidtFromDet})
\begin{equation}
    \det(\widetilde{\rho}_m)>p_m^2\,4\,\alpha_0\alpha_1\beta_0\beta_1\quad\forall\,m.
    \label{eqn:proofXZ-1}
\end{equation}
Since $\det(\widetilde{\rho}_m)=\alpha_0\alpha_1\beta_0\beta_1\,
\vert\det(\widehat{u}_m)\vert^2$, the summation over $m$ of the square root of
Eq.~(\ref{eqn:proofXZ-1}) yields
\begin{equation}
    \sum_{m=1}^n \vert \det(\widehat{u}_m)\vert >2.
    \label{eqn:proofXZ-2}
\end{equation}
But the concurrence of a (normalized) state is smaller
or equal than one, hence $2\,|\det(\widehat{u}_m)|\leq\Vert u_m\Vert^2$.
Moreover, taking the trace of the completeness relation for the operators $E_m$ implies
$\sum_{m=1}^n \Vert u_m\Vert^2=4$. Therefore $\sum_{m=1}^n
\vert \det(\widehat{u}_m)\vert\leq2$, which is in contradiction with
Eq.~(\ref{eqn:proofXZ-2}) and concludes the proof.
\end{Pf}

% -----------------------------------------------------------------------------
% SCP
% -----------------------------------------------------------------------------
\subsubsection{Maximizing the SCP}
\label{sssec:1d-1r-scp}
The following result gives the maximum value of the SCP for one entanglement
swapping step:\\

% Result
% .............................................................................
\begin{Res}\label{res:zz-basis}
    Best SCP strategy for one repeater
  \begin{quote}
    The measurement that maximizes $\SCP$   for a one-repeater configuration
    is the Bell measurement in the ZZ basis, and
    \begin{equation}
        \SCP_{\max}=\SCP_{\text{ZZ}}=2\,\min\{\alpha_1,\beta_1\}.
        \label{eqn:1repeater-SZZ}
    \end{equation}
  \end{quote}
\end{Res}

% Proof
% .............................................................................
\begin{Pf}
    Two kinds of outcomes appear when performing a Bell measurement in the
    computational basis: two of the outcome probabilities are equal to $p_{\max}$,
    while the other two are equal to $p_{\min}$. Putting these values into Eq.~(\ref{eqn:projM})
    one finds the corresponding smallest Schmidt coefficients:
    \begin{equation}
        \lambda(p_{\max})=\frac{\alpha_1\beta_1}{2\,p_{\max}},\quad
        \lambda(p_{\min})=\frac{\min\{\alpha_0\beta_1,\,\alpha_1\beta_0\}}{2\,p_{\min}},
        \label{eqn:lambdasZZ}
    \end{equation}
    whence $\SCP_{\text{ZZ}}=2\,\min\{\alpha_1,\beta_1\}$.
    Consider now that we are allowed to perform some arbitrary unitary not only
    on $BC$, but on $ABC$. We are in presence of a bipartite system, and results of
    majorization theory apply: the SCP of this system is at most $2\,\beta_1$. A
    similar construction for qubits $B$, $C$ and $D$ tells us that the SCP
    is at most $2\,\alpha_1$, so that the final SCP cannot exceed twice the minimum
    of $\alpha_1$ and $\beta_1$.
\end{Pf}

% Remark
% -----------------------------------------------------------------------------
\paragraph*{Remark}
Setting $\alpha=\beta$, one sees that the SCP does not decrease after one
entanglement swapping; this is the ``conserved entanglement'' described in \cite{BVK99}.

% =============================================================================
% 2 Repeaters
% =============================================================================
\subsection{Two Repeaters}
\label{ssec:1d-2r}
We consider a system of three states on which we perform two consecutive
entanglement swappings, as shown in Fig.~\ref{fig:notation1D}c, and we describe the
measurements that maximize the three figures of merit.

% -----------------------------------------------------------------------------
% C and WCE
% -----------------------------------------------------------------------------
\subsubsection{Maximizing the Concurrence and the WCE}
\label{sssec:1d-2r-cwce}
The maximization of these two figures of merit is trivial for a two-repeater
configuration once one knows the results for the one-repeater system. First,
any Bell measurement maximizes the average concurrence
of the results of the two measurements. This will be generalized and proved for
any number of repeaters in \S\ref{sssec:1d-infinite-cwce}. Then, in order to maximize the WCE,
we simply have to perform XZ measurements at both repeaters. In fact, if we perform any other measurement on the first
repeater, then at least one resulting state $\varphi_m$ will be less entangled than the
XZ results, and this reflects on the WCE of the second measurement (which has
to be a Bell measurement in the XZ basis from Result~\ref{res:xz-basis}).

% -----------------------------------------------------------------------------
% SCP
% -----------------------------------------------------------------------------
\subsubsection{Maximizing the SCP}
\label{sssec:1d-2r-scp}
After the first measurement, we get four resulting states $\ket{\varphi_m}$ with
probabilities $p_m$. From Result~\ref{res:zz-basis} we know that for any
outcome, the second measurement has to be done in the ZZ basis.
Hence, we have to find the first measurement $M$ that maximizes
\begin{equation}
    \SCP_M^{(2)}(\alpha_0,\,\beta_0,\,\gamma_0)= 2\,\sum_{m} p_m\,\min\{\varphi_{m,1},\,\gamma_1\}.
    \label{eqn:2repeaters-bestS}
\end{equation}

We first maximize this quantity over the set of Bell measurements (which, as we will
see, leads to the best strategy for a large range of entangled states $\alpha$,
$\beta$ and $\gamma$), and then we present some numerical results showing that non-Bell
measurements sometimes provide better results.\\

\begin{figure}
    \psfrag{p}[][]{$p$}
    \psfrag{f}[][]{$f(p)$}
    \psfrag{g}[][]{$g(p)$}
    \psfrag{h}[][]{$h(p)$}
    \psfrag{ps}[][]{$p^*$}
    \psfrag{hs}[][]{$2\gamma_1p^*$}
  \begin{center}
      \includegraphics[height=3.02cm]{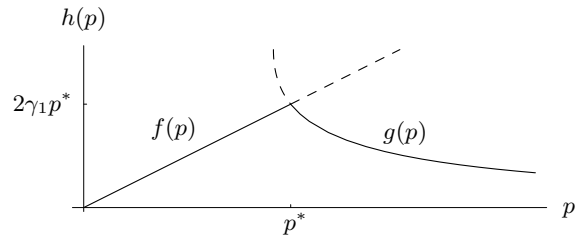}
      \caption{Representation of the function $h(p)=\min\{f(p),\,g(p)\}$ governing the SCP after Bell measurements in a two-repeater configuration.}
      \label{fig:2repeaters-h}
  \end{center}
\end{figure}

% Bell measurements
% -----------------------------------------------------------------------------
\paragraph*{Bell Measurements}
We fix the states $\alpha$, $\beta$ and $\gamma$ and we consider the SCP as
a function of the outcome probabilities only:
\begin{equation}
    \SCP(\{p_m\}) = \sum_m \min\{f(p_m),\,g(p_m)\}\equiv \sum_m h(p_m),
    \label{eqn:2repeat-scpBell}
\end{equation}
where $f(p) = 2\gamma_1p$ and $g(p) = p-\sqrt{p^2-\alpha_0\alpha_1\beta_0\beta_1}$.
One can show that $g'(p)<0$ and $g''(p)>0$ $\forall\,p\in[p_{\min},\,p_{\max}]$.
A typical plot of $h(p)$ is shown in Fig.~\ref{fig:2repeaters-h}, and the value $p^*$ at which
the functions $f$ and $g$ cross each other is
\begin{equation}
    p^* = \frac{1}{2}\sqrt{\frac{\alpha_0\alpha_1\beta_0\beta_1}{\gamma_0\gamma_1}}.
    \label{eqn:2repeaters-pstar}
\end{equation}

% 2repeaters-h.eps ::::::::::::::::::::
% :::::::::::::::::::::::::::::::::::::

It is sufficient to maximize the function over the possible probability distributions,
since Result~\ref{res:bell-prob} insures the existence of a Bell measurement leading to
this optimal distribution; we recall that the probabilities have to be
chosen in the interval $[p_{\min},p_{\max}]$. Let us give two necessary conditions
that have to be satisfied by the best probability distribution (they can be proven rigorously,
but a look at Fig.~\ref{fig:2repeaters-h} may be clearer):
\begin{itemize}
    \item if the set $\{p_m\}$ maximizes $\SCP$, then all probabilities lie either to the
        left of $p^*$ or to its right. In fact, suppose for example that
        $p_1+2\varepsilon<p^*<p_2-2\varepsilon$,
        and choose $\tilde{p}_1=p_1+\varepsilon$ and $\tilde{p}_2=p_2-\varepsilon$
        (with $0<\epsilon\ll1$ as it should be). The constraints on these new probabilities
        are clearly satisfied if it was the case before, and a better SCP has been found.
    \item if $p_1$ and $p_2$ are such that $p^*+2\varepsilon<p_1\leq p_2<p_{\max}-2\varepsilon$,
        then the choice $\tilde{p}_1=p_1-\varepsilon$ and $\tilde{p}_2=p_2+\varepsilon$
        gives rise to a strictly greater SCP (this comes from the convexity of $g$).
\end{itemize}

It is now simple to maximize the SCP of two repeaters, and one sees that the value
$p^*$, with respect to $p_{\min}$ and $p_{\max}$, plays a crucial role in the
choice of the best probability distribution. In fact, we have to distinguish
four distinct cases; see results in Tab.~\ref{tab:2repeaters}. We notice
that ZZ measurements lead to the maximum SCP whenever $p^*\leq p_{\min}$,
while the XZ ones are the best strategy for $p^*\geq1/4$.
So far, we have maximized the SCP for two repeaters supposing that
the first measurement was to be done on the states $\alpha$ and $\beta$.
But what happens if we start from the right side? It appears that the maximum SCP depends, in general,
on the order of the measurements and that performing the first measurement where
the states are more entangled yields better results.\\

% tab:2repeaters ::::::::::::::::::::::
\begin{table}
    \begin{center}
    \begin{tabular}{c|c}
        value of $p^*$ & $\{p_m\}$ maximizing $\SCP^{(2)}$\\
        \hline
        \phantom{$^\big|$}
        $p^*\leq p_{\min}$ & $\{p_{\min},\,p_{\min},\,p_{\max},\,p_{\max}\}$\\
        \phantom{$^\big|$}
        $p_{\min}\leq p^*\leq(1-p_{\max})/3$ & $\{p^*,\,p^*,\,p_{\max},\,1-2p^*-p_{\max}\}$\\
        \phantom{$^\big|$}
        $(1-p_{\max})/3\leq p^*\leq1/4$ & $\{p^*,\,p^*,\,p^*,\,1-3p^*\}$\\
        \phantom{$^\big|$}
        $p^*\geq1/4$ & $\{1/4,\,1/4,\,1/4,\,1/4\}$
    \end{tabular}
    \caption{Maximization of $\SCP^{(2)}$ over Bell measurements, see text for details.}
    \label{tab:2repeaters}
    \end{center}
\end{table}
% :::::::::::::::::::::::::::::::::::::

% Numerical Results
% -----------------------------------------------------------------------------
\paragraph*{General Measurements (Numerical Results)}
The question is to check if some non-Bell measurements yield a better SCP
than the results of the last paragraph. Since the concurrence
of the states used for entanglement swapping can now take any value between
0 and 1, we cannot consider $\SCP$ as a function of the outcome probabilities
only. But for a fixed concurrence $C<1$ one sees that
\[
    \bar{g}(C,\,p)\equiv p-\sqrt{p^2-\alpha_0\alpha_1\beta_0\beta_1\,C^2}<g(p)\quad\forall\,p.
\]

\begin{figure}
    \psfrag{x}[][]{$\alpha_0$}
    \psfrag{h}[][]{$\SCP_{\max}^{(2)}(\alpha_0,\beta_0,\gamma_0)$}
    \psfrag{0.1}[][]{$0.1$}
    \psfrag{0.2}[][]{$0.2$}
    \psfrag{0.3}[][]{$0.3$}
    \psfrag{0.4}[][]{$0.4$}
    \psfrag{0.5}[][]{$0.5$}
    \psfrag{0.69}[][]{$a_1$}
    \psfrag{0.94}[][]{$a_2$}
    \psfrag{1}[][]{$1$}
  \begin{center}
      \includegraphics[height=3.3cm]{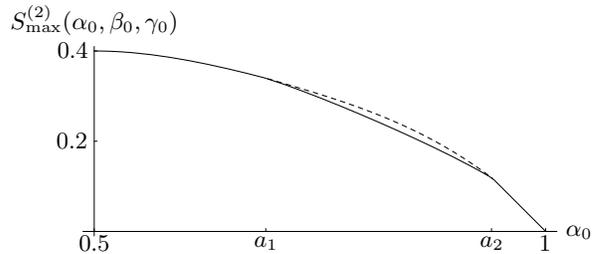}
      \caption{SCP for a system of two repeaters, with $\beta_0=\gamma_0=0.7$.
        Numerical results (dashed line) show that there exists a better strategy
        than Bell measurements (solid line) for $\alpha_0\in[a_1,\,a_2]$. The
        values $a_1$ and $a_2$ are such that $p^*(a_1)=(1-p_{\max}(a_1))/3$
        and $p^*(a_2)=p_{\min}(a_2)$.}
      \label{fig:2repeaters-nonBell}
  \end{center}
\end{figure}

Writing the corresponding variables of non-Bell measurements with a bar, we have
that $\bar{p}^*<p^*$ and $\bar{g}(C,\,\bar{p}^*) < g(p^*)$.
Therefore, one can check that Bell measurements are indeed the best ones,
except, possibly, when $p_{\min}\leq p^*\leq(1-p_{\max})/3$. The key fact
about Bell measurements in that case is that we cannot chose three outcome probabilities
to lie on $p^*$, since the fourth one would be greater than $p_{\max}$.
But the range of possible outcome probabilities depends on the concurrence: for
example, from Eq.~(\ref{eqn:projM-p}) and for $C(u_m)=0$, we have that $\bar{p}_m\in[\alpha_1
\beta_1,\,\alpha_0\beta_0]$, or more generally
\begin{equation}
    \bar{p}_m \in [\bar{p}_{\max},\,\bar{p}_{\min}] \supseteq [p_{\max},\,p_{\min}].
\end{equation}
Hence, and this is confirmed by numerical results,
a better strategy is to perform a measurement such that three outcomes
probabilities are equal to $\bar{p}^*$, and that the concurrences of the states are
the largest ones satisfying $\bar{p}_{\max}=1-3\,\bar{p}^*$. Our numerical evidence shows that Bell measurements do not always maximize the SCP, see
Fig.~\ref{fig:2repeaters-nonBell}.

% 2repeaters-nonBell.eps ::::::::::::::
% :::::::::::::::::::::::::::::::::::::

% =============================================================================
% 1D Chain
% =============================================================================
\subsection{Large  1D Chains}
\label{ssec:1d-infinite}
We consider the system of Fig.~\ref{fig:notation1D}a that consists of $N$ repeaters
joining $N+1$ states. For simplicity, we choose the states $\alpha_i$ as
being identical: $\ket{\alpha_i}=\ket{\varphi}\;\forall\,i$.
We show in this section which strategies yield the optimal
solution for the concurrence and the WCE, and for the SCP we give an upper bound
to its maximum value and some results for XZ and ZZ measurements.

% -----------------------------------------------------------------------------
% C and WCE
% -----------------------------------------------------------------------------
\subsubsection{Maximizing the Concurrence and the WCE}
\label{sssec:1d-infinite-cwce}
A direct generalization of Eq.~(\ref{eqn:projM-C}) for $N$ repeaters yields for
the concurrence \cite{VMC04}:
\begin{equation}
    C^{(N)} = \sum_{\{m_i\}}2\,|\det\left(X_{\{m_i\}}\right)|,
    \label{eqn:avgC}
\end{equation}
where $X_{\{m_i\}}=\widehat{\varphi}\,\widehat{u}_{m_1}\,\widehat{\varphi}\ldots \widehat{u}_{m_N}\,\widehat{\varphi}$,
and the states $\ket{u_{m_i}}$ are associated with the measurement result
$m_i$ of the $i$-th repeater. Then the maximization of $C^{(N)}$ reads
\begin{multline}
    \underset{M}{\max}\big\{C^{(N)}\big\}
        = |\det(\widehat{\varphi})|^{N+1}\ \max_M\Big\{\sum_{\{m_i\}}2^{N+2}\\
            \Big|\det\big(\widehat{\Phi}\,\widehat{u}_{m_1}\,\widehat{\Phi}
            \ldots \widehat{u}_{m_N}\,\widehat{\Phi}\big)\Big|\Big\}
        = |2\,\det(\widehat{\varphi})|^{N+1},
        \label{eqn:maxavgC}
\end{multline}
where $\widehat{\Phi}=\id_2/\sqrt{2}$ corresponds to a maximally entangled
state. For states $\varphi$ which are not perfect singlets, the concurrence
decreases exponentially with $N$:
\begin{equation}
    C_{\max}^{(N)}\sim(4\,\varphi_0\varphi_1)^{N/2},\quad N\gg1.
    \label{eqn:limC-1D}
\end{equation}

The same arguments as for the systems of one or two repeaters hold for the WCE, so that XZ Bell measurements
have to be performed on each repeater in order to maximize it.

% -----------------------------------------------------------------------------
% SCP
% -----------------------------------------------------------------------------
\subsubsection{Maximizing the SCP}
\label{sssec:1d-infinite-scp}
A similar formula as Eq.~(\ref{eqn:avgC}) for the average SCP is
\begin{equation}
    \SCP^{(N)} = \sum_{\{m_i\}}2\,\min\left\{\eig\left(X_{{m_i}}^{\phantom{\dagger}}X_{{m_i}}^{\dagger}\right)\right\}.
    \label{eqn:avgSCP}
\end{equation}
Contrary to the maximization of the concurrence, we cannot find here such an easy
way to calculate the maximum value of $\SCP$, but we already can say a few words
about the SCP for a 1D chain with a large number of repeaters:
\begin{itemize}
    \item Since $\SCP$ is always smaller than or equal to $C$, it is
        upper bounded by
        \begin{equation}
            \SCP_{\max}^{(N)}\lesssim(4\,\varphi_0\varphi_1)^{N/2}
            \label{eqn:1d-upperbound}
        \end{equation}
    \item After $N\gg1$ measurements, the entanglement of the resulting states
        is expected to be, in average, very small, so that the SCP and the concurrence
        could be related by $\SCP\approx C^2$. Hence we may have the asymptotic
        behavior $\SCP^{(N)}\sim (4\,\varphi_0\varphi_1)^N$.
\end{itemize}

Even if we do not have the protocol that maximizes the SCP,
we present here three specific strategies, as the results are instructive.
The first and simplest one consists in trying to convert each state into a singlet, and then
to establish a perfect connection between the end-qubits of the chain. In the second strategy
we perform XZ measurements at all stations, and from \S\ref{sssec:1d-1r-cwce} we
know that all resulting states have the same amount of entanglement. We
indeed find the exponentional decay of the SCP related to the one of the concurrence. Finally, in App.~\ref{sec:app-scpzz}, we derive
the explicit formula for ZZ measurements on a chain of any number of repeaters,
which yields an decay of the SCP which is quite close
to the upper bound given in Eq.~(\ref{eqn:1d-upperbound}). The asymptotic behaviors
are summarized in Tab.~\ref{tab:1d-asympt}.

% tab:1d-asympt :::::::::::::::::::::::
\begin{table}
    \begin{center}
    \begin{tabular}{c|ccccc}
       	& CS & & XZ & & ZZ\\
        \hline
        \phantom{$^\big|$}$S^{(N)}$ &
        $(2\,\varphi_1)^N$ & $<$ &
        $(4\,\varphi_0\varphi_1)^N$ & $<$ &
        $\frac{1}{\sqrt{N}}(4\,\varphi_0\varphi_1)^{N/2}$
    \end{tabular}
    \caption{Asymptotic behavior of the SCP for a 1D chain of $N\gg1$ repeaters. Three specific
    measurement protocols are studied: conversion of all states into singlets (CS), XZ and ZZ measurements.}
    \label{tab:1d-asympt}
    \end{center}
\end{table}
% :::::::::::::::::::::::::::::::::::::

% #############################################################################
% 2D
% #############################################################################
\section{The Simplest 2D Network: a Square}
\label{sec:2d}

% square.eps ::::::::::::::::::::::::::
\begin{figure}
    \psfrag{a}[][]{$\alpha$}	\psfrag{b}[][]{$\beta$}
    \psfrag{j}[][]{$\varphi$}	\psfrag{f}[][]{$\psi$}
    \psfrag{M1}[][]{$M_1$}
    \psfrag{M2}[][]{$M_2$}
  \begin{center}
      \includegraphics[height=1.8cm]{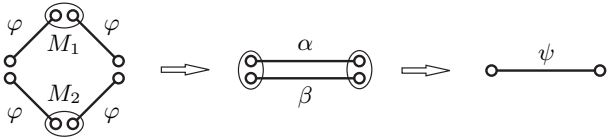}
      \caption{Operations on a square to obtain an entangled pair on the diagonal: first two
      measurements, then distillation of the resulting states $\alpha$ and $\beta$.}
      \label{fig:square}
  \end{center}
\end{figure}
% :::::::::::::::::::::::::::::::::::::

We study in this section a square made of four identically entangled states, see Fig.~\ref{fig:square}.
This is clearly one of the simplest possible 2D networks.
The operations we perform consist of three steps: a first measurement
$M_1$ yielding some outcome $\alpha$, then a measurement $M_2$ depending on $\alpha$
and giving another state $\beta$, and finally a distillation of these two states to get
a final state $\psi$. The goal is of course to get $\psi$ as entangled as possible,
given the states $\varphi$.

% -----------------------------------------------------------------------------
% Distillation
% -----------------------------------------------------------------------------
\setcounter{subsubsection}{0}
\subsubsection{Distillation}
\label{sssec:2d-square-pur}
Majorization theory \cite{NV01} tells us how entangled the state
$\psi$ can be. Without loss of generality we choose $\alpha_0\geq\beta_0$
and the majorization criterion reads
\begin{equation}
    \left(\alpha_0\beta_0,\,\alpha_0\beta_1,\,\alpha_1\beta_0,\,\alpha_1\beta_1\right)
    \prec
    \left(\psi_0,\,\psi_1,\,0,\,0\right),
\end{equation}
whose only non-trivial inequality is $\alpha_0\beta_0 \leq \psi_0$.
Since we are looking for a state $\psi$ that is as entangled as possible, we
know its greatest Schmidt coefficient:
\begin{equation}
    \psi_0=\max\Big\{\frac{1}{2},\,\alpha_0\beta_0\Big\}.
    \label{eqn:square-psi0}
\end{equation}

% -----------------------------------------------------------------------------
% Figures of Merit
% -----------------------------------------------------------------------------
\subsubsection{Maximizing the Figures of Merit}
\label{sssec:2d-square-merit}
Arguments used for 1D networks still hold here, so that one has to perform
Bell measurements and XZ measurements to maximize the concurrence and the WCE respectively.
It is worth pointing out that a perfect singlet $\psi$ can be established
with probability one after two XZ measurements and a distillation if $\varphi$
satisfies
\begin{equation}
    \varphi_0 \leq \varphi_0^*=\frac{1+\sqrt{1-\sqrt{2\left(\sqrt{2}-1\right)}}}{2}
    \approx 0.65.
\label{eqn:0.65}
\end{equation}

Thus we consider that $\varphi$ is less entangled that $\varphi^*$
since we already know how to get a singlet for
$\varphi_0 \leq \varphi_0^*$. We proceed in two steps for maximizing the SCP: we first look at the
subproblem of maximization over the measurements $M_2$ for a given
outcome $\alpha$, and then we provide some numerical results for the whole
square.\\

% Second Measurement
% -----------------------------------------------------------------------------
\paragraph*{Second Measurement}
We first notice that a singlet can be obtained by a XZ measurement with probability
one if $\alpha_0\leq\alpha_0^{\star}\equiv\left(1+\sqrt{1-(4\,\varphi_0\varphi_1)^2}\right)^{-1}$.
Then, labeling by $m$ the resulting states $\beta$ of the measurement $M_2$, we can write the function to be maximized as
\begin{align}
    \SCP_M^{\triangle}
        &= 2\sum_m p_m\left(1-\max\Big\{\frac{1}{2},\,\alpha_0\beta_{m,0}\Big\}
            \right)\notag\\
        &= 2\,\alpha_1+\alpha_0\,2\sum_m p_m\,\min\Big\{\beta_{m,1},\,
        \frac{\alpha_0-\alpha_1}{2\,\alpha_0}\Big\}\notag\\
        &\equiv \SCP_{\max}^{(0)}(\alpha_0)+\alpha_0\,\SCP_M^{(2)}\Big(\varphi_0,\,
            \varphi_0,\,\frac{1}{2\,\alpha_0}\Big),
\end{align}
so that all results of Sect.~\ref{ssec:1d-2r} can be applied.
The three quantities $p_{\min}$, $p_{\max}$ and $p^*$ used in that section are now
$p_{\min} = \varphi_0\varphi_1$, $p_{\max} = (\varphi_0^2+\varphi_1^2)/2$ and
$p^* = \varphi_0\varphi_1\,\alpha_0/(\alpha_0-\alpha_1)$.
Since $p^*$ is greater than $p_{\min}$ for all states $\alpha$ and $\varphi$, it follows that
$\SCP_{\max}^{\triangle}$ is reached by Bell measurements except when
$p^*\in\,\left]p_{\min},\,(1-p_{\max})/3\right[$.\\

% First Measurement
% -----------------------------------------------------------------------------
\paragraph*{First Measurement}

% square-h.eps ::::::::::::::::::::::::
\begin{figure}
    \psfrag{1}[][]{$1$}
    \psfrag{p}[][]{$p$}
    \psfrag{hp}[][]{$h(p)$}
    \psfrag{2.14}[][]{$p_{\min}^{\phantom{\star}}$}
    \psfrag{2.26}[][]{$p^{\star}$}
    \psfrag{2.5}[][]{$0.25^{\phantom{\star}}$}
    \psfrag{2.86}[][]{$p_{\max}^{\phantom{\star}}$}
  \begin{center}
    \includegraphics[height=3cm]{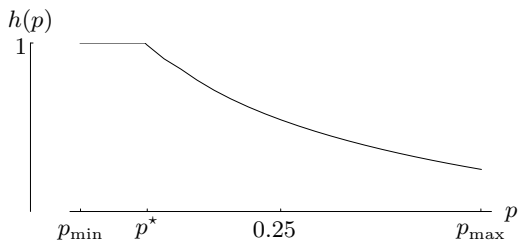}
    \caption{Typical plot of the function $h(p)$ governing the SCP of the square.}
    \label{fig:square-h}
  \end{center}
\end{figure}
% :::::::::::::::::::::::::::::::::::::

The function to maximize over the measurements $M_1$ is
\begin{equation}
    \SCP^{\square}_{M_1}=\sum_m p_m\,
        \SCP_{\max}^{\triangle}(\alpha_{m,0},\,\varphi_0,\,\varphi_0).
\end{equation}

\begin{figure}
    \psfrag{6.5}[][]{\small{$\varphi_0^*$}}
    \psfrag{6.64}[][]{\small{$\varphi_0^{\star}$}}
    \psfrag{0.9}[][]{$0.9$}
    \psfrag{1}[][]{$1$}
    \psfrag{x}[][]{$\varphi_0$}
    \psfrag{h}[][]{$\SCP_{\max}^{\square}$}
  \begin{center}
    \includegraphics[height=3cm]{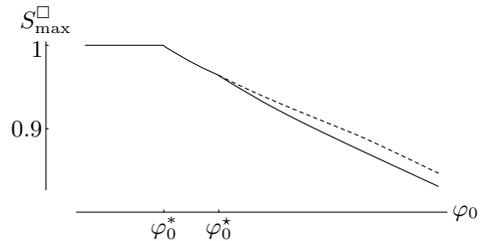}
    \caption{SCP for a square made of four states $\varphi$. Numerical results
      (dashed line) show that Bell measurements (solid line) do not lead to the
      optimal solution for $\varphi_0>\varphi^{\star}\approx0.664$.}
    \label{fig:square-scp}
  \end{center}
\end{figure}

For Bell measurements, since the Schmidt coefficient $\alpha_{m,0}$ depends on
$p_m$ only, we can write $\SCP^{\square}_{M_1} = \sum_m h(p_m)$. Here we make a slightly abuse of notation, since we again use $h(p)$, as in \S\ref{sssec:1d-2r-scp}. Actually, the shape and properties of the function $h(p)$ discussed here and in \S\ref{sssec:1d-2r-scp} are very similar. Therefore, all arguments used in that section for the
maximization of the SCP apply here, too. The plot of $h(p)$ is shown in Fig.~\ref{fig:square-h}. The quantity that corresponds to $p^*$ is
now written $p^{\star}$ and its value is
\[
    p^{\star}=\frac{\varphi_0\varphi_1}{2\sqrt{\alpha_0^{\star}\alpha_1^{\star}}},
\]
where $\alpha_1^{\star}\equiv1-\alpha_0^{\star}$. With these definitions, one
can check that for all $\varphi_0$ greater than $\varphi_0^*$, we have $p_{\min}\leq p^{\star}
(\varphi_0) \leq 1/4$ and that $p^{\star}\rightarrow p_{\min}$ when $\varphi_0\rightarrow1$,
whence the best measurements for nearly unentangled states $\varphi$ are
the ZZ ones. As for the system of two repeaters, performing Bell measurements
is not the best choice when it is not possible to get three of the four
outcome probabilities to be equal to $p^{\star}$ (but this possible when
$\varphi_0\leq\varphi_0^{\star}\approx0.664$, see Fig.~\ref{fig:square-scp}).
 Finally, we summarize the results
in Tab.~\ref{tab:square}, and the similarity with Tab.~\ref{tab:2repeaters} is immediate.

% square-scp.eps ::::::::::::::::::::::
% :::::::::::::::::::::::::::::::::::::

% tab:square ::::::::::::::::::::::::::
\begin{table}
    \begin{center}
    \begin{tabular}{c|c}
        value of $p^*$ & $\{p_m\}$ maximizing $\SCP^{\square}$\\
        \hline
        \phantom{$^\big|$}
        $\varphi_0\rightarrow1$ & $\{p_{\min},\,p_{\min},\,p_{\max},\,p_{\max}\}$\\
        \phantom{$^\big|$}
        $\varphi_0\geq\varphi_0^{\star}$ & $\{p^{\star},\,p^{\star},\,p_{\max},\,1-2p^{\star}-p_{\max}\}$\\
        \phantom{$^\big|$}
        $\varphi_0^*\leq\varphi_0\leq\varphi_0^{\star}$ & $\{p^{\star},\,p^{\star},\,p^{\star},\,1-3p^{\star}\}$\\
        \phantom{$^\big|$}
        $\varphi_0\leq \varphi_0^*$ & $\{1/4,\,1/4,\,1/4,\,1/4\}$
    \end{tabular}
    \caption{Maximization of $\SCP^{\square}$ over Bell measurements, see text for details.}
    \label{tab:square}
    \end{center}
\end{table}
% :::::::::::::::::::::::::::::::::::::

% #############################################################################
% Hierarchical Lattices
% #############################################################################
\section{Hierarchical Lattices}
\label{sec:hier}
In this section we will apply directly the results of the previous sections to study establishment of entanglement over large scales in lattices with hierarchical geometry. These are lattices that iterate certain geometric structures, so that at each level of iteration the number of nodes or the number of neighbors changes. Unfortunately, we do not know how to find optimal strategies for such lattices; we restrict our considerations to show that one can establish perfect entanglement in a finite number of steps at some iteration level. This perfect entanglement can be swapped further to the lowest levels of iteration, \ie to the largest
scales, which can be considered as the largest geometrical distances. One should stress that classical percolation strategies work for hierarchical lattices as well as for 2D or 3D ones. Nevertheless, every percolation strategy relies on conversion of all bonds to singlets with a certain probability $p$ and then establishing
a perfect entanglement between two nodes on the large scale with a probability  $\theta^2(p)$. The latter formula expresses the fact that both nodes have to belong to the percolating cluster, which happens for each of them independently with probability $\theta(p)$ \cite{Grimmett}. This probability is always smaller than one, except for the trivial case of $p=1$.

% =============================================================================
% Diamond Lattice
% =============================================================================
\subsection{``Diamond'' Lattice}

\begin{figure}
	\psfrag{A}[][]{$A$}			\psfrag{B}[][]{$B$}
	\psfrag{C}[][]{$C$}			\psfrag{D}[][]{$D$}
  \begin{center}
    \includegraphics[height=2.2cm]{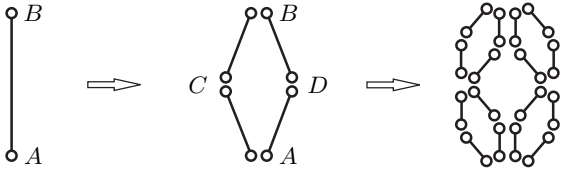}
    \caption{The diamond lattice is formed by iterating the following operation: a
    single bond (two qubits and one entangled state) is replaced by four bonds
    forming a diamond shape (four pairs of qubits and four entangled states).
    After $K$ iterations, the nodes $A, B, C, D$ have $2^K$ links, the nodes on
    the next level $2^{K-1}$ links, \etc}
    \label{fig:diam}
  \end{center}
\end{figure}

We start considering the so-called ``diamond'' lattice, which is obtained by iterating the operation presented in Fig.~\ref{fig:diam}, in which
a single bond (two qubits and one entangled state) is replaced by four bonds forming a diamond shape (four pairs of qubits and four entangled states).
We prove that for sufficiently large initial entanglement, one can establish perfect entanglement on large scales (\ie on some lower levels of iteration) in finite number of steps.

% diamond.eps :::::::::::::::::::::::::
% :::::::::::::::::::::::::::::::::::::

\begin{figure}
	\psfrag{0}[][]{\footnotesize{0}}			\psfrag{0.2}[][]{\footnotesize{0.2}}
	\psfrag{0.4}[][]{\footnotesize{0.4}}	\psfrag{0.6}[][]{\footnotesize{0.6}}
	\psfrag{0.8}[][]{\footnotesize{0.8}}	\psfrag{1}[][]{\footnotesize{1}}
	\psfrag{E}[][]{$E_{th}$}
	\psfrag{th}[][]{}
	\psfrag{Initial Entanglement E}[][]{\footnotesize{Initial entanglement $E$}}
	\psfrag{Final Entanglement E'}[][]{\footnotesize{Final entanglement $E'$}}
  \begin{center}
    \includegraphics[height=5.4cm]{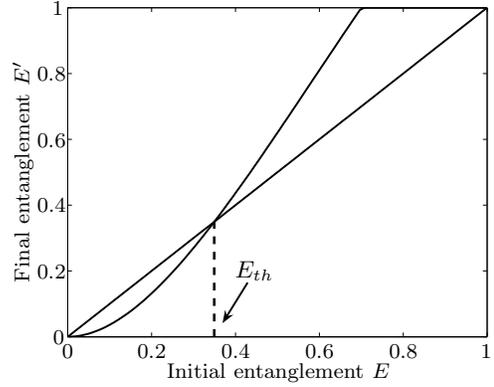}
    \caption{Recursion relating $E$ on the higher level of lattice iteration to
    $E'$ at the lower level of iteration  in the  diamond lattice. Each
    iteration consists of the following steps: (i) WCE and (ii) the two
    resulting two-qubit states are transformed with probability one into a
    two-qubit state of the same SCP.}
    \label{fig:rec-diam}
  \end{center}
\end{figure}

We assume that the lattice is formed by very many iterations, and that all bonds correspond to entangled states $\ket{\varphi}=\varphi_0\ket{00}+ \varphi_1\ket{11}$. Our aim is to perform measurements in a recursive way and demonstrate the for sufficiently small $\varphi_0$ it is possible to establish perfect entanglement on the lowest level of the iteration hierarchy, \ie between the ``parent'' nodes $A$ and $B$. In order to keep the form of the network unchanged during the recursive measurement we will apply the WCE strategy to the nodes analogue to $C$ and $D$, staring from the highest (last) iteration level. After applying WCE we obtain with probability 1 a pair of entangled states $\ket{\psi}=\psi_0\ket{00}+ \psi_1\ket{11}$,
with $ \psi_0=(1 + \sqrt{1-16\varphi_0^2\varphi_1^2})/2$. This pair can then be distilled with probability 1 to a new two-qubit entangled state $\ket{\varphi'}$, see Eq.~(\ref{eqn:square-psi0}):
\[
	\varphi'_0= \max\left\{\frac{1}{2},\,\frac{1}{4}\Big(1 + \sqrt{1-16\varphi_0^2\varphi_1^2}\Big)^2\right\}.
\]
Denoting now the SCP by $E=2\varphi_1$, we rewrite the recursion as
\[
	E'= 2(1-\psi_0^2)= 1 + (2-E)^2E^2/2 -\sqrt{1-(2-E)^2E^2}.
\]
This recursion (see Fig.~\ref{fig:rec-diam}) has one nontrivial unstable fixed point $E_{th}$, and two trivial stable fixed points $\tilde E=0$ and $\tilde E=1$. The latter is achieved in a finite number of steps provided the initial $E>E_{th}\approx0.349$. Note that $E_{th}$ is strictly smaller than $E^*= 2(1-\varphi_0^*)$ from Eq.~(\ref{eqn:0.65}).  For $E\geq E^*$, $E'$ is equal to 1, \ie the singlet is achieved in one step.

% rec-diam.eps ::::::::::::::::::::::::
% :::::::::::::::::::::::::::::::::::::

% =============================================================================
% Tree Lattice
% =============================================================================
\subsection{``Tree'' Lattice}
\begin{figure}
\begin{center}
	\psfrag{(a)}[][]{(a)}						\psfrag{(b)}[][]{(b)}
	\psfrag{E}[][]{$E$}							\psfrag{E0}[][]{$E_0$}
	\psfrag{E1}[][]{$E_{\romone}$}	\psfrag{E2}[][]{$E_{\romtwo}$}
	\psfrag{F}[][]{$E'$}
  \includegraphics[width=6.6cm]{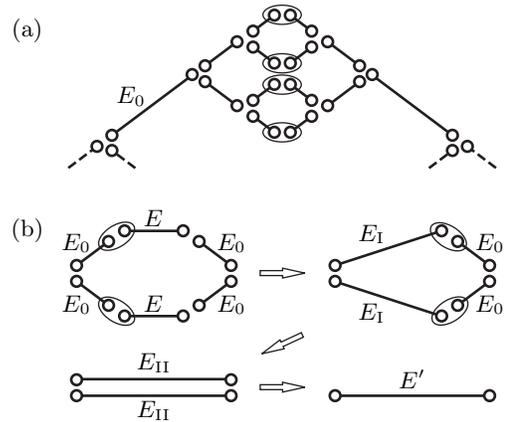}
  \caption{(a) Tree configuration; (b) The nodes in the middle perform
  WCE. This creates two two-qubit states between
  the neighboring nodes. These states are transformed into a two-qubit state of the same SCP. The process is iterated until a perfect singlet is established
  between the two ends of the tree.}
  \label{fig:tree1}
\end{center}
\end{figure}

Similar results hold for the simplest possible ``tree'' lattice: a double Cayley tree lattice
with branching factor two (see Fig.~\ref{fig:tree1}a).

% tree1.eps :::::::::::::::::::::::::::
% :::::::::::::::::::::::::::::::::::::

Let us denote the initial SCP of all bonds by $E_0$. The strategy is depicted in Fig.~\ref{fig:tree1}: the nodes in the middle of the tree perform the WCE. This prepares two two-qubit states between the neighboring nodes with the entanglement, measured by the SCP, equal to $E_1$. These two states are then converted with probability 1 into a two-qubit state with $E=\min\{1,\,2(1-(1-E_1/2)^2)\}$, which will undergo recursive transformations (Fig.~\ref{fig:tree1}b). We perform then the WCE on one of the 3 connected bonds, and obtain $E_{\romone}=1- \sqrt{1-E_0(2-E_0)E(2-E)}$. Then, the WCE is applied to
the remaining  pair of bonds yielding $E_{\romtwo}=1- \sqrt{1-E_0(2-E_0)E_{\romone}(2-E_{\romone})}$.
Finally, the optimal singlet conversion is applied to the pair of $E_{\romtwo}$ bonds obtained from the two different but neighbouring branches of the tree, yielding $E'=\min\{1,\,2(1-(1-E_{\romtwo}/2)^2)\}$.  The recursion relations can be rewritten as
\begin {equation}
	E'=F(E;E_0).
	\label{eqn:rec-tree}
\end{equation}
This recursion depends explicitely on $E_0$. It is easy to see that since the WCE does not increase the SCP,
the recursion (\ref{eqn:rec-tree}) will have: i) only one trivial stable fixed point $\tilde E=0$ if $E_0<E_{th}$;
ii) three fixed points otherwise: stable $0$, unstable $\tilde E$ and stable $1$ otherwise. Here if we start with $E\ge E_{th}$ we will
end up in one step with $E'=1$. The threshold value is obtained by solving the equation
$1=2(1-(1-(1- \sqrt{1-E_0^2(2-E_0)^2})/2)^2)$, and is given by
\begin{equation}
    E_{th}=1-\sqrt{1-\sqrt{2(\sqrt 2-1)}}\approx0.7.
\end{equation}

% #############################################################################
% Genuine 2D Lattices
% #############################################################################
\section{Genuine 2D Lattices}
\label{sec:2D}

In this section we consider genuine 2D lattices when the number of nodes is big. On the one hand, we apply
the methods and observations of the previous sections to these large lattices. On the other hand, we reconsider
the various variants of the methods employing classical and quantum percolation techniques.

% =============================================================================
% Centipede in Square Lattice
% =============================================================================
\subsection{``Centipede'' in Square Lattice}

\begin{figure}
  \begin{center}
  	\psfrag{(a)}[][]{(a)}						\psfrag{(b)}[][]{(b)}
		\psfrag{E}[][]{$E$}							\psfrag{E0}[][]{$E_0$}
		\psfrag{E1}[][]{$E_{\romone}$}	\psfrag{E2}[][]{$E_{\romtwo}$}
		\psfrag{F}[][]{$E'$}
    \includegraphics[height=6.5cm]{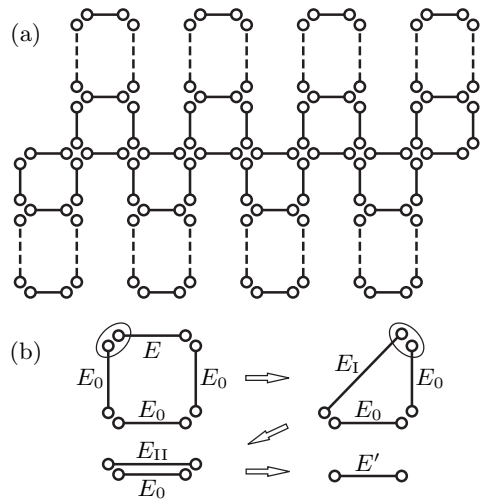}
    \caption{(a) ``Centipede'' with its ``legs'' and ``spine''; (b) recursive measurement scheme; note that the method
    can be equally well applied also in higher dimensions.}
    \label{fig:centi}
  \end{center}
\end{figure}

As another example of the power of recursive measurement methods of the previous section,  we consider
a wide strip of a 2D square lattice and the ``centipede'' figure within it (see Fig.~\ref{fig:centi}a). Let us denote the initial entanglement as $E_0$, and the entanglement at the end bond of a ``leg'' by $E$. We then apply the following measurement scheme  to the ends of each of the legs of the centipede, see also
Fig.~\ref{fig:centi}: i) We apply the WCE to $E_0$ and $E$, replacing these two bonds by one with $E_{\romone}=1- \sqrt{1-E_0(2-E_0)E(2-E)}$; (ii)  we repeat it with the other vertical bond obtaining thus a pair of states at the horizontal end of the leg: one with entanglement $E_0$ and the other with $E_{\romtwo}=1- \sqrt{1-E_0(2-E_0)E_{\romone}(2-E_{\romone})}$; (iii) the resulting pair is then distilled with probability 1 to a two-qubit state
with $E'=\min\{1,\,2(1-(1-E_0/2)(1-E_{\romtwo}/2))\}=F(E;E_0)$. This situation is somewhat similar to the case
of the tree lattice from the previous section, but not completely.
The recurrence relation depends explicitly on $E_0$ and has always a nontrivial stable fixed point
$\tilde E>E_0$.  This fixed point, however, is strictly smaller than 1, when $E_0$ is small.
In the first case, although we do concentrate more entanglement along the ``spine'' we still have to face the problem that the spine is a 1D network, and will exhibit
exponential decrease of probability of establishing the perfect entanglement for large distances \cite{ACL07}.
On the other hand, the stable fixed point  is simply $\tilde E=1$,
provided $E_0$ is large enough.
In this  case a perfect singlet is achieved in a finite number of steps,
and the singlets from all legs can be concentrated at the spine of the centipede with probability 1. Obviously, all that implies that the width of the strip of the 2D lattice (equal to twice the length
of the centipede leg) can be finite: it must be just larger than the number of steps necessary to get a perfect singlet.

% centi.eps :::::::::::::::::::::::::::
% :::::::::::::::::::::::::::::::::::::

The condition for the threshold value of $E_0$ is easy to derive: we have to put $E=1$ in the above recurrence and solve
$1=2(1-(1-E_{th}/2)(1+\sqrt{1-E_{th}^2(2-E_{th})^2})/2),$
which gives $E_{th}\approx0.649$.

% centi-rec.eps :::::::::::::::::::::::
\begin{figure}
	\psfrag{0}[][]{\footnotesize{0}}			\psfrag{0.2}[][]{\footnotesize{0.2}}
	\psfrag{0.4}[][]{\footnotesize{0.4}}	\psfrag{0.6}[][]{\footnotesize{0.6}}
	\psfrag{0.8}[][]{\footnotesize{0.8}}	\psfrag{1}[][]{\footnotesize{1}}
	\psfrag{E}[][]{$E$}
	\psfrag{>E}[][]{ $\;>\negmedspace E$}
	\psfrag{=E}[][]{ $\;=\negmedspace E$}
	\psfrag{<E}[][]{ $\;<\negmedspace E$}
	\psfrag{th}[][]{ \textit{\footnotesize{ \;th}}}
	\psfrag{Initial Entanglement E}[][]{\footnotesize{Initial entanglement $E$}}
	\psfrag{Final Entanglement E'}[][]{\footnotesize{Final entanglement $E'$}}
  \includegraphics[height=5.4cm]{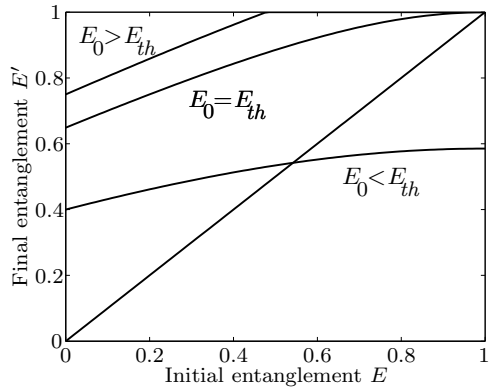}
  \caption{Recursion relation for the centipede lattice. Only when the entanglement $E_0$ is larger than a threshold $E_{th}$, a trivial stable point at $E=1$ appears.}
  \label{fig:centi-rec}
\end{figure}
% :::::::::::::::::::::::::::::::::::::

% =============================================================================
% Percolation Strategies
% =============================================================================
\subsection{Percolation Strategies}
In Ref.~\cite{ACL07} we have pointed out that one possible strategy for entanglement distribution is to convert locally all bonds with probability $p$ into singlets and then perform entanglement swapping. This strategy can then be linked to classical percolation theory, so all the known results of this field can be applied to our quantum scenario.
The natural question is whether the thresholds defined by classical percolation theory are optimal or entanglement percolation represents a related but different theoretical problem where new bounds have to be obtained. This of course equivalent to determine whether the measurement
strategy based on SCP is optimal in the asymptotic regime. Here, we construct several examples that go beyond the classical percolation picture, proving that the classical entanglement percolation strategy is not optimal. The key ingredient for the construction of these examples is the measurement strategy previously obtained for the 1D one-repeater configuration that maximizes the SCP.

% tab:percolation :::::::::::::::::::::
\begin{table}
    \begin{center}
    \begin{tabular}{l@{\quad}|@{\quad}ll}
				lattice & $p_c$ & \\
				\hline
				triangular
					\phantom{$^{\big|}$}
										& $p_c^{\vartriangle}$ 	& $= 2\,\sin(\pi/18)\approx 0.347$\\
				square			& $p_c^{\Box}$					& $= 0.5$\\
				honeycomb		& $p_c^{\hexagon}$			&	$= 1-2\,\sin(\pi/18)\approx 0.653$
			\end{tabular}
    \caption{Classical (bond) percolation thresholds $p_c$ for some regular lattices.}
    \label{tab:percolation}
    \end{center}
\end{table}
% :::::::::::::::::::::::::::::::::::::

% -----------------------------------------------------------------------------
% Honeycomb Lattice with Doubled Bonds
% -----------------------------------------------------------------------------
\subsubsection{Honeycomb Lattice with Doubled Bonds}
\begin{figure}
	\psfrag{(a)}[][]{(a)}
	\psfrag{(b)}[][]{(b)}
	\begin{center}
	  \includegraphics[width=7.2cm]{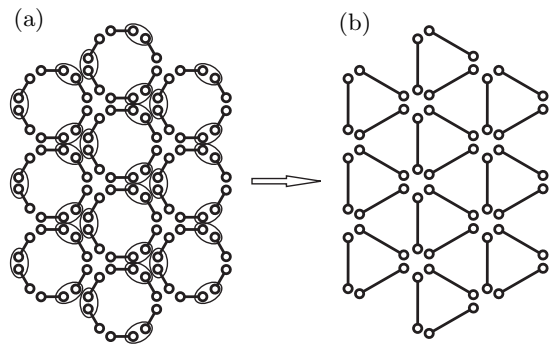}
	  \caption{Each node is connected by two copies of the same two-qubit state
	  $\ket{\varphi}$. The nodes marked in (a) perform the measurement optimal
	  according to the SCP. A triangular lattice (b) is obtained where the SCP
	  is the same as for the state $\ket{\varphi}$. Classical entanglement
	  percolation is now possible in the new lattice.}
	  \label{fig:hontr}
	\end{center}
\end{figure}

The first example (already discussed in Ref.~\cite{ACL07})
considers a honeycomb lattice where each node is connected by two copies of the same two-qubit state $\ket{\varphi}$, see Fig.~\ref{fig:hontr}a.\\

The simplest strategy consists in trying to convert all bonds of the doubled honeycomb lattice into singlets,
and then applying the classical entanglement percolation. The percolation threshold of this doubled lattice is not difficult to calculate:
at the critical point, the probability that, at each edge, at least one conversion is successful has to be equal to the
percolation threshold of the simple honeycomb lattice; if both conversions are successful we simply discard one pair.
We thus have $p_c^{\hexagon}=1-(1-2\varphi_1)^2$, hence the percolation threshold is (see Tab.~\ref{tab:percolation})
\begin{equation}
	2\varphi_1 = 1-\sqrt{2\,\sin\left(\frac{\pi}{18}\right)}\approx 0.411.
\end{equation}
We define the classical entanglement percolation strategy as i) converting in the best possible way \textit{all} bonds shared by two parties into \textit{one} singlet and ii) applying the entanglement percolation. If, as above, the Schmidt coefficients of the two-qubit state are $\varphi_0\geq\varphi_1$, the SCP of $\ket{\varphi}^{\otimes 2}$ is given by $p_{ok}=2(1-\varphi_0^2)$. We choose this conversion probability to be equal to the percolation threshold for the honeycomb lattice and get
\begin{equation}
	2\varphi_1 = 2\left(1-\sqrt{\frac{1}{2}+\sin\left(\frac{\pi}{18} \right)}\right) \approx 0.358.
\end{equation}
We now show that another strategy yields a better percolation threshold: some of the nodes, see Fig.~\ref{fig:hontr}a, perform the optimal strategy for the SCP, mapping the honeycomb lattice into a triangular lattice, as shown in Fig.~\ref{fig:hontr}b. What is important is that the SCP for the new bonds is exactly the same as for the initial state $\ket{\varphi}$, that is $2\varphi_1$. We choose it to be equal to $p_c^{\vartriangle}$, so that
\begin{equation}
	2\varphi_1=2\,\sin\left(\frac{\pi}{18}\right)\approx 0.347,
\end{equation}
which proves that the classical entanglement strategy is not the best one.

% hontr.eps :::::::::::::::::::::::::::
% :::::::::::::::::::::::::::::::::::::

% -----------------------------------------------------------------------------
% Asymmetric Triangular Lattice
% -----------------------------------------------------------------------------
\subsubsection{Asymmetric Triangular Lattice}
\begin{figure}
	\psfrag{(a)}[][]{(a)}		\psfrag{(b)}[][]{(b)}
	\begin{center}
	  \includegraphics[width=7.6cm]{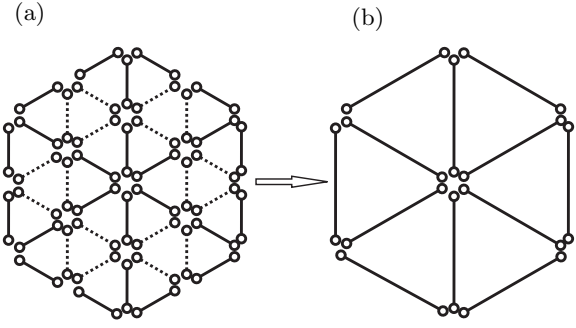}
	  \caption{The triangular lattice consists of two different entangled states
	  $\ket{\varphi}$ and $\ket{\tilde\varphi}$ for the solid and dashed lines,
	  respectively. The less entangled states, $\ket{\tilde\varphi}$, are discarded
	  and some of the nodes perform the optimal measurement according to the SCP. A
	  new triangular lattice is obtained, governed by the SCP of $\ket{\varphi}$.}
	  \label{fig:asymtr}
	\end{center}
\end{figure}

The second type of examples, although less symmetric, is generic and has a totally different character than the previous one. For simplicity, we show the argument in the case of a triangular lattice, but the same reasoning can be applied to other geometries. Consider the triangular lattice of Fig.~\ref{fig:asymtr}a. Solid lines correspond to two-qubit pure states $\ket{\varphi}$ while dashed lines correspond to states $\ket{\tilde\varphi}$ that are less entangled, \ie $\tilde\varphi_0>\varphi_0$. We choose the first state such that $p_{ok}=2\varphi_1$ satisfies $p_c^\vartriangle< p_{ok}<\sqrt{p_c^\vartriangle}$. If $\ket{\tilde\varphi}=\ket{\varphi}$, the classical entanglement percolation strategy works. However, we choose this second less entangled state such that its SCP is small enough to make the classical entanglement percolation impossible. This state always exists. Indeed, note that when $\varphi_1\to 0$, these states can simply be removed from the lattice, and classical entanglement percolation fails because of $p_{ok}^2<p_c^\vartriangle $. It is now rather straightforward to construct a successful entanglement percolation strategy: the state $\ket{\tilde\varphi}$ are discarded and the optimal strategy for the one-repeater configuration and the SCP is performed. The lattice is then mapped into a new triangular lattice keeping the conversion probability of the first, more entangled, state, see Fig.~\ref{fig:asymtr}b. Classical entanglement percolation can now be applied to this new lattice, since $p_{ok}>p_c^\vartriangle$.

% asymtr.eps ::::::::::::::::::::::::::
% :::::::::::::::::::::::::::::::::::::

% =============================================================================
% Doubling the Square Lattice
% =============================================================================
\subsection{Doubling the Square Lattice}
\label{ssec:2D-FKG}
\begin{figure}
	\psfrag{(a)}[][]{(a)}		\psfrag{(b)}[][]{(b)}
	\psfrag{A}[][]{$A$}			\psfrag{B}[][]{$A'$}
	\begin{center}
  	\includegraphics[width=7.6cm]{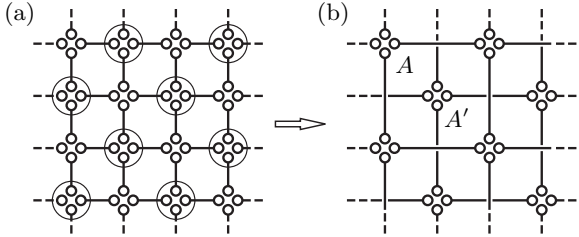}
  	\caption{(a) Measurements necessary to double the square lattice: the marked
  	nodes apply the optimal one-repeater transformation along the vertical and
  	horizontal directions. (b) Resulting pairs of disjoint square lattices with
  	lattice constant doubled; we want to establish perfect entanglement between
  	any two neighboring points $A$, $A'$ versus $B$ and $B'$. $A$ and $A'$ ($B$
  	and $B'$) are neighbors but belong to different lattices.}
  	\label{fig:double}
	\end{center}
\end{figure}

The final example deals with a square lattice and has yet another character.
Here we replace every second pair of horizontal bond by a single one using the optimal SPC strategy, which as we know from Sec.~\ref{sec:1d}
does not change the SCP on average, replacing, however, pure states by a known mixture.
The same is done with every second pair of vertical bonds. In effect we replace the original square lattice by two disjoint lattices with the lattice constant twice bigger than the original one, but
the same SCP (see Fig.~\ref{fig:double}). Now we consider the following problem: we are interested in establishing entanglement between any of the two neighboring nodes $A$, $A'$ and $B$, and $B'$ at large distances.

% disjlatt.eps ::::::::::::::::::::::::
% :::::::::::::::::::::::::::::::::::::

In the case of doubled lattices the calculation is simple: the pairs $(A,B)$ and $(A',B')$ belong to two
disjoint lattices, and the probability that, say, $(A,B)$ belongs to the percolating cluster is equal asymptotically to $\theta^2(p)$.
The probability that at least one of the pairs belongs to the percolating cluster is thus
\begin{equation}
	P_{\text{double}}=2\,\theta^2-\theta^4=\theta^2\;(2-\theta^2).
\end{equation}
This probability has to be compared with the probability that at least one of the pairs $(A,B)$, $(A',B)$, $(A,B')$ or $(A',B')$ belongs to the percolating cluster in the original square lattice. The latter probability is asymptotically
$\pi^2$ where $\pi$ is the probability that $A $ or $A'$ (or equivalently $B$ or $B'$) belongs to the percolating cluster $\cal C$.
Thus we have
\begin{align*}
  \pi &= P[A\,\text{or}\, A' \in {\cal C}]\\
   		&= P[A \in{\cal C}] + P[A' \in {\cal C}] - P[A,A'\in {\cal C}].
\end{align*}
In order to estimate the last term in the above expression, we use the FKG inequality
(\cite{FKG71}, see also \cite{Grimmett}). To state it, we first define an event described in terms of a
percolation configuration to be {\it increasing} if it has the property that, once it
holds for a certain bond configuration, it holds for all configurations obtained by
adding bonds to the initial one.  FKG inequality says that any two such events are
positively correlated. The events $\{A \in {\cal C}\}$ and $\{A \connected A'\}$
(``$A$ and $A'$ are connected by a path of maximally entangled bonds'') are clearly
increasing and, since their intersection is the event $\{A,A' \in {\cal C}\}$, it
follows that
\begin{equation}
	P[A,A'\in{\cal C}] \geq P[A\in{\cal C}]\,P[A \connected A'].
\end{equation}
Denoting $P[A \connected A']$ by $\tau$, we thus have
$\pi^2\leq \theta^2\;(2 - \tau)^2$.
Therefore, doubling the square lattice is a better strategy than the classical percolation,
\ie $\pi^2\leq P_{\text{double}}$, whenever
\begin{equation}
	(2 - \tau)^2 \leq 2 - \theta^2.
\end{equation}
We believe that this inequality is fulfilled for all $p$,
and we show it when $p$ is just above the percolation threshold $p_c^{\Box}=0.5$, so that
$\theta$ tends to zero. To this aim we have to demonstrate that $2-\tau\leq\sqrt{2}$.
We may try to estimate $\tau$ from below by considering the six shortest  
trajectories connecting $A$ and $A'$: the most direct ones, two 2-edge  
paths, and the two pairs of 4-edge paths around the adjacent squares. One finds
\[
	\tau > 2\,(p^2+2p^4-2p^5) -  (p^2+2p^4-2p^5)^2.
\]
Unfortunately, for $p=p_c^{\Box}$ this estimate is too small, since it  
gives only  $2-\tau < 1.473\ldots$ and adding further paths becomes  
then technically tedious. We therefore turn to the standard
numerical Monte Carlo method, generating the shortest paths  
automatically while the longer ones are generated using the Monte Carlo  
sampling. For $p>p_c^{\Box}$ the convergence is exponential: if we plot a  
subsequent estimate of $\tau$ as a function of the maximum cluster  
size allowed in the Monte Carlo sampling, it approaches the final   
value exponentially fast for large clusters. As expected, the  
convergence is algebraic at $p=p_c^{\Box}$: the estimate of  
$\tau$ approaches its final value as a power of the cluster size. A  
power law fit and a comparison with the values just above the  
percolation threshold give with a very good accuracy $\tau \simeq 0.687$ and
hence $2-\tau \simeq 1.313 <\sqrt{2}$, \quoded

This is yet another result which does not have a classical analogue, showing how
quantum mechanical measurements allow to increase percolation probability.

% #############################################################################
% Conclusion
% #############################################################################
\section{Conclusion}
\label{sec:conclu}
In this paper, we have considered the problem of entanglement percolation through
pure-state quantum networks. We have first focused our investigations on small quantum networks.
Even for these particularly simple systems, interesting and unexpected
properties have been pointed out. One of the main result is the description of a Bell measurement by its outcome probabilities only (Result~\ref{res:bell-prob}).
This has allowed us to maximize the different figures of merit introduced at
the beginning of the paper. We have shown, then, that Bell
measurements do not yield in general the optimal protocol, even for a chain consisting of only two repeaters.

The results for small lattices have later be used as building blocks for entanglement percolation protocols in asymptotically large lattices. We have provided several examples illustrating some of the properties characterizing these lattices: recursive relations, classical entanglement percolation examples of lattices were quantum effects allow going beyond classical percolation.

In general, little is still known about the problem of entanglement percolation, that is, the distribution of entanglement through quantum networks. In the pure-state case, it would be interesting to derive lower bounds to the amount of entanglement between the nodes such that entanglement percolation is possible. The main question, however, is to extend these results to the mixed-state scenario, providing examples of entanglement percolation protocols for lattice with mixed-state bonds.

% *****************************************************************************
% ACKNOWLEDGMENTS, BIBLIOGRAPHY & APPENDIX
% *****************************************************************************
\begin{acknowledgments}
	We thank John Lapeyre who kindly supplied us with a high-precision numerical value of
	$P[A \connected A']$ in \S\ref{ssec:2D-FKG}.
  Much of this work was supported by the QCCC program, part of the Elite Network of Bavaria (ENB).
  We also acknowledge support from
  the cluster of excellence Munich Centre for Advanced Photonics (MAP),
  the Deutsche Forschungsgemeinschaft,
  the EU IP programs `SCALA' and `QAP',
  the European Science Foundation PESC QUDEDIS,
  and the MEC (Spanish Government) under contracts FIS 2005-04627, FIS 2004-05639 and Consolider QOIT.
\end{acknowledgments}

\appendix
\numberwithin{equation}{section}

% #############################################################################
% PROOF OF RESULT 1
% #############################################################################
\section{Proof of Result~\ref{res:bell-prob}}
\label{sec:app-proof-result}

We prove here that there always exists a Bell measurement which yields outcome
probabilities $p_m$ equal to $x_m$, when these values add up to one
and lie in the interval $[p_{\min},\,p_{\max}]$.\\

% Proof
% .............................................................................
\begin{Pf}  \textit{(By contradiction).}
    Let us write $\{\mu_m\}$ the four states of the Bell measurement
    in the magic basis.
    Because the matrix $(\mu_{m,i})$ is orthogonal, the conditions
    on $x$ are clearly necessary. In fact, we know from Eq.~(\ref{eqn:notation-p}) that
		$p_m = p_{\min}\,k_m + p_{\max}\,(1-k_m)$ with $k_m\in[0,1]$.
    One of the four equations of the system $\{p_m\}=\{x_m\}$ will
    be dependent of the other three: if we can find three orthogonal
    vectors $\mu_m$ such that $p_m=x_m$ for, say, $m=1,2,3$, then the
    fourth one is fixed (up to a sign) with, obviously, $p_4=x_4$.
    Let us write these three states $\mu_m$ as
    \begin{multline*}
        \mu_m = \big(\sqrt{k_m}\,\cos(\theta_m),\,
                         \sqrt{k_m}\,\sin(\theta_m),\\
                         \sqrt{1-k_m}\,\cos(\omega_m),\,
                         \sqrt{1-k_m}\,\sin(\omega_m)\big),
    \end{multline*}
    where $k_m = (p_{\max}-x_m)/(p_{\max}-p_{\min})$.
    By construction, these vectors are normalized and satisfy $p_m=x_m$. We now
    have to prove that there always exist some angles $\theta_m$ and $\omega_m$ such
    that these three vectors are orthogonal. Without loss of generality we
    order the $k$'s such that $1\geq k_1\geq k_2\geq k_3\geq k_4 \geq 0$. Since
    the probabilities add up to 1 and that $p_{\min}+p_{\max}=0.5$ we have
    \begin{equation}
        k_1+k_2+k_3+k_4=2.
        \label{eqn:proof-sumk}
    \end{equation}
    Introducing the notations $k_m'\equiv 1-k_m$, $\theta_a\equiv\theta_1-\theta_2$,
    $\theta_b\equiv\theta_1-\theta_3$, $\omega_a\equiv\omega_1-\omega_2$,
    $\omega_b\equiv\omega_1-\omega_3$ and using the identity $\cos(x)\cos(y)+\sin(x)\sin(y)=\cos(x-y)$,
    the conditions of orthogonality read
    \begin{equation}
    \left\{
        \begin{array}{r@{\:=\:}l}
        0 & \sqrt{k_1k_2}\,\cos(\theta_a)+\sqrt{k_1'k_2'}\,\cos(\omega_a)\\
        0 & \sqrt{k_1k_3}\,\cos(\theta_b)+\sqrt{k_1'k_3'}\,\cos(\omega_b)\\
        0 & \sqrt{k_2k_3}\,\cos(\theta_a-\theta_b)+\sqrt{k_2'k_3'}\,\cos(\omega_a-\omega_b)
        \end{array}
    \right.
    \label{eqn:bellproof-orthogonal}
    \end{equation}
    The cases $k_m=0$ or $k_m=1$ for some $m$ can be trivially solved, so
    we consider $k_m\neq0$ and $k_m'\neq0$. We have four parameters
    $\theta_{a,b}$ and $\omega_{a,b}$ which can be freely chosen in the interval
    $[0,\,\pi]$, but the two inequalities $\sqrt{k_1k_2}\geq\sqrt{k_1'k_2'}$ and
    $\sqrt{k_1k_3}\geq\sqrt{k_1'k_3'}$ impose the constraints
    $\theta_a\in[\theta_a^*,\,\pi-\theta_a^*]$ and $\quad\theta_b\in[\theta_b^*,\,\pi-\theta_b^*]$,
    with $\theta_{a,b}^*\in[0,\,\frac{\pi}{2}]$ such that
    $\sqrt{k_1k_2}\,\cos(\theta_a^*)=\sqrt{k_1'k_2'}$ and
    $\sqrt{k_1k_3}\,\cos(\theta_b^*)=\sqrt{k_1'k_3'}$.
    Thus $\cos(\omega_a-\omega_b)\in[-1,\,1]$ and
    $\cos(\theta_a-\theta_b)\in[-\cos(\theta_a^*+\theta_b^*),\,1]$. Then,
    one can verify that there always exists at least one solution of Eq.~(\ref{eqn:bellproof-orthogonal}),
    except when $-\sqrt{k_2k_3}\,\cos(\theta_a^*+\theta_b^*)>\sqrt{k_2'k_3'}$, what
    never happens. In fact, suppose that this last inequality holds and rewrite
    it in terms of $k_1$, $k_2$ and $k_3$ only. After some tedious algebra and
    using some trigonometric identities, one finds
    that the inequality $k_1+k_2+k_3>2$ holds, but this is in contradiction
    with Eq.~(\ref{eqn:proof-sumk}), which concludes the proof.
\end{Pf}

% #############################################################################
% SCP ZZ 1D
% #############################################################################
\section{SCP of ZZ Measurements on a 1D Chain}
\label{sec:app-scpzz}

% tree.eps  :::::::::::::::::::::::::::
\begin{figure}
	\psfrag{n=0}[][]{$n=0$}
  \psfrag{n=1}[][]{$n=1$}
  \psfrag{0}[][]{$0$} \psfrag{1}[][]{$1$} \psfrag{2}[][]{$2$}
  \psfrag{p+0}[][]{$\scriptstyle p_+(0)$} \psfrag{p-0}[][]{$\scriptstyle p_-(0)$}
  \psfrag{p+1}[][]{$\scriptstyle p_+(1)$} \psfrag{p-1}[][]{$\scriptstyle p_-(1)$}
  \begin{center}
    \includegraphics[height=3.2cm]{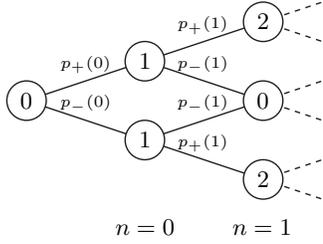}
    \caption{``Tree view'' of the labels $m$ and their corresponding probabilities
    after the first two measurements. For symmetry, we choose the root of this tree
    corresponding to $n=-1$.}
    \label{fig:tree}
	\end{center}
\end{figure}
% :::::::::::::::::::::::::::::::::::::

Even if the number of outcomes grows exponentially with the number of repeaters,
one can keep track of all of them in an efficient way. In fact, after any number $n\leq N$ of entanglement swappings in the ZZ basis, any
possible resulting state has the form (up to local unitaries)
\[
\ket{m} \equiv \frac{1}{\sqrt{\varphi_0^m+\varphi_1^m}}
    \left(\sqrt{\varphi_0^m}\ket{00}+\sqrt{\varphi_1^m}\ket{11}\right),\quad
    m\in\mathbb{N}.
\]
We prove this by induction on $n$, the case $n=0$ corresponding to the initial state $m=1$.
Suppose that the result holds and that we got the state $\ket{m}$ after $n<N$ measurements.
It is easy to show from Eq.~(\ref{eqn:lambdasZZ}) that an entanglement swapping
in the ZZ basis on $\ket{m}\otimes\ket{\varphi}$ is described by
\begin{equation}
    \ket{m} \mapsto \begin{cases}
        \ket{m+1}       & \text{with probability}\;p_+(m)\\
        \ket{\,|m-1|\,} &   \text{with probability}\;p_-(m),
    \end{cases}
\end{equation}
where $p_+(m)=(\varphi_0^{m+1}+\varphi_1^{m+1})/(\varphi_0^m+\varphi_1^m)$ and $p_-(m)=1-p_+(m)$, \quoded\\

The first step to calculate the SCP of this protocol is to compute its variation after a ZZ measurement.
Considering that the set $\{(p_i,\ket{m_i}),\,i=1,\ldots,l\}$ describes all
the resulting states of $n$ measurements, and writing $\lambda_{\pm}(m)$ the smallest
Schmidt coefficient of $\ket{m\pm1}$, the new SCP reads
\begin{align}
    \SCP_{\text{ZZ}}^{(n+1)}
        &= \sum_{i=1}^lp_i\,2\big(p_+(m_i)\,\lambda_+(m_i)+p_-(m_i)\,\lambda_-(m_i)\big)\notag\\
        &= \SCP_{\text{ZZ}}^{(n)}-(\varphi_0-\varphi_1)\:p(m=0,n),
\label{eqn:scpZZ-1}
\end{align}
where $p(m=0,n)$ stands for the probability of getting the state $\ket{m=0}$ after
$n$ measurements. Since this probability is not zero for $n$ odd only, it results
that the SCP decreases for $n$ even only.
We have now to calculate the probability $p(m=0,n)$ of getting a singlet after $n$ measurements:
it is the weighted sum over all possible paths $\Gamma$ that
go from the root node $m=0$ to the node $m=0$ at position $n$ in the tree drawn in Fig.~\ref{fig:tree}.
We notice that the weight, $w$, depends on $n$ only and not on $\Gamma$. This is indeed the fact since
$p_+(m)\,p_-(m+1) = \varphi_0\,\varphi_1$ for all $m$ and because we have to go
up in the tree as many times as we have to go down. Thus, for $n$ odd we have
$w(n) = (\varphi_0\varphi_1)^{(n+1)/2}$ and using basic combinatorial analysis one finds that
$p(m=0,n) = (\varphi_0\varphi_1)^{k} \binom{2k}{k}$, with $k=\frac{1}{2}(n+1)\in\mathbb{N}$.
Finally, denoting by $[x]$ the integer part of $x$, the general expression of the SCP for a chain of $N$
repeaters reads
\begin{equation}
    \SCP_{\text{ZZ}}^{(N)} = 1 - (\varphi_0-\varphi_1)\,\sum_{k=0}^{[N/2]}
        (\varphi_0\varphi_1)^{k} \binom{2k}{k}.
    \label{eqn:scpZZ}
\end{equation}
%It is clear, by construction, that this value is positive and tends to 0 as $N$ goes to infinity. This can be explicitly shown by performing the change of variable
%$\varphi_0=\left(1+\sqrt{1-y}\right)/2$ where $y\in[0,1[$, such that the sum becomes the Taylor series of order $[N/2]$ of $1/\sqrt{1-y}$:
%\[
%    \SCP_{\text{ZZ}}^{(N)} = 1 - \sqrt{1-y}\,\sum_{k=0}^{[N/2]} \frac{1}{4^k}
%        \binom{2k}{k}\,y^k.
%\]

% #############################################################################
% BIBLIOGRAPHY
% #############################################################################
\bibliography{main}

\end{document}